\documentclass[8.5pt,twoside,twocolumn]{article}
\usepackage[super,sort&compress,comma]{natbib}
\usepackage[version=3]{mhchem}
\usepackage{times,mathptmx}
\usepackage{sectsty}
\usepackage{balance}

\usepackage{graphicx} 
\usepackage{amssymb}
\usepackage{lastpage}
\usepackage[format=plain,justification=raggedright,singlelinecheck=false,font=small,labelfont=bf,labelsep=space]{caption}
\usepackage{fancyhdr}

\usepackage{color}

\begin{document}

\thispagestyle{plain}
\fancypagestyle{plain}{
\fancyhead[L]{\includegraphics[height=8pt]{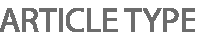}}
\fancyhead[C]{\hspace{-1cm}\includegraphics[height=20pt]{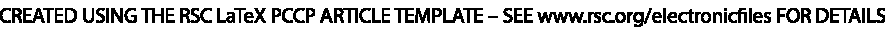}}
\fancyhead[R]{\includegraphics[height=10pt]{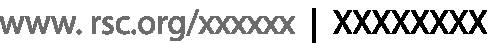}\vspace{-0.2cm}}
\renewcommand{\headrulewidth}{1pt}}
\renewcommand{\thefootnote}{\fnsymbol{footnote}}
\renewcommand\footnoterule{\vspace*{1pt}%
\hrule width 3.4in height 0.4pt \vspace*{5pt}}
\setcounter{secnumdepth}{5}

\def\<{\langle}
\def\>{\rangle}

\makeatletter
\def\subsubsection{\@startsection{subsubsection}{3}{10pt}{-1.25ex plus -1ex minus -.1ex}{0ex plus 0ex}{\normalsize\bf}}
\def\paragraph{\@startsection{paragraph}{4}{10pt}{-1.25ex plus -1ex minus -.1ex}{0ex plus 0ex}{\normalsize\textit}}
\renewcommand\@biblabel[1]{#1}
\renewcommand\@makefntext[1]%
{\noindent\makebox[0pt][r]{\@thefnmark\,}#1}
\makeatother
\renewcommand{\figurename}{\small{Fig.}~}
\sectionfont{\large}
\subsectionfont{\normalsize}

\fancyfoot{}
\fancyfoot[LO,RE]{\vspace{-7pt}\includegraphics[height=9pt]{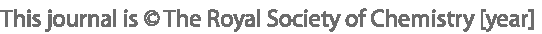}}
\fancyfoot[CO]{\vspace{-7.2pt}\hspace{12.2cm}\includegraphics{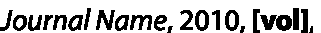}}
\fancyfoot[CE]{\vspace{-7.5pt}\hspace{-13.5cm}\includegraphics{RF}}
\fancyfoot[RO]{\footnotesize{\sffamily{1--\pageref{LastPage} ~\textbar  \hspace{2pt}\thepage}}}
\fancyfoot[LE]{\footnotesize{\sffamily{\thepage~\textbar\hspace{3.45cm} 1--\pageref{LastPage}}}}
\fancyhead{}
\renewcommand{\headrulewidth}{1pt}
\renewcommand{\footrulewidth}{1pt}
\setlength{\arrayrulewidth}{1pt}
\setlength{\columnsep}{6.5mm}
\setlength\bibsep{1pt}

\twocolumn[
  \begin{@twocolumnfalse}

\noindent\LARGE{\textbf{Universal behaviour of the glass and the jamming transitions in finite 
dimensions for hard spheres}
\vspace{0.6cm}

\noindent\large{%
Antonio Coniglio\textit{$^{a\ddag}$}},
Massimo Pica Ciamarra\textit{$^{b,a}$},
and 
Tomaso Aste\textit{$^{c}$}
\vspace{0.5cm}
}

\noindent\textit{\small{\textbf{Received Xth XXXXXXXXXX 20XX, Accepted Xth XXXXXXXXX 20XX\newline
First published on the web Xth XXXXXXXXXX 200X}}}

\noindent \textbf{\small{DOI: 10.1039/b000000x}}
\vspace{0.6cm}

\noindent \normalsize{ 
We investigate the glass and the jamming transitions of hard spheres in
finite dimensions $d$, through a revised cell theory, that combines the
free volume and the Random First Order Theory (RFOT). 
Recent results show that in infinite dimension the ideal glass transition 
and jamming transitions are distinct, while
based on our theory we argue that they indeed coincide for finite $d$.  
As a consequence, jamming results into a percolation
transition described by RFOT, with a static length diverging with
exponent $\nu=2/d$, which we verify through finite size scaling,
and standard critical exponents $\alpha = 0$, $\beta = 0$ and $\gamma = 2$
independent on $d$.
} \vspace{0.5cm}
 \end{@twocolumnfalse}
]

\footnotetext{\textit{$^{a}$~CNR-SPIN, Dipartimento di Fisica, Universit\`a ``Federico II'', Napoli, Via Cintia, 80126 Napoli, Italy.}}
\footnotetext{\textit{$^{b}$~Division of Physics and Applied Physics, School of Physical and Mathematical Sciences, Nanyang Technological University, Singapore}}
\footnotetext{\textit{$^{c}$~Department of Computer Science, University College London, Gower Street, London, WC1E 6BT, UK.}}
\footnotetext{\textit{$^{\ddag}$~corresponding author mail: coniglio@na.infn.it}}


\section{Introduction}
\begin{figure}[t!]
\includegraphics*[scale=0.2]{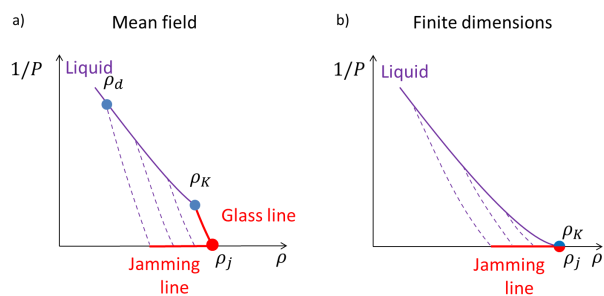}
\caption{
Schematic phase diagram for a hard sphere system.
The full line represents the `equilibrium' equation of state 
within the 
metastable fluid phase, while dashed lines represent
out-of-equilibrium compression protocols. The infinite pressure limit of these 
lines
define the jamming line. 
Panel a illustrates mean field results
elaborated from Fig.~4 of Ref.~\cite{kurchan2009}, while panel b
illustrates the finite dimensional scenario~\cite{Liu10}
supported by our results.
Here the ideal glass transition critical density $\rho_K$ 
coincides with an ideal jamming critical density $\rho_j$, 
both being located at the end of an equilibrium line.
\label{fig:schematic}
}
\end{figure}

The transition between a fluid to an amorphous solid phase
is common to many disordered systems, such as
molecular liquids, colloids, granular materials and foams,
and its understanding is one of the major problems
in condensed matter.
In a seminal paper~\cite{nagel1998},
an universal jamming phase diagram was proposed 
to unify the transition of structural arrest of different systems, including the 
glass
and the jamming transitions.
However, within a mean field or the infinite dimensionality limit
a distinction between the glass transition of liquids and the jamming transition
of granular materials was posed
in an analytical study of frictionless hard sphere particles~\cite{PZ2010, 
kurchan2009,charbonneau2014,Kurchan13,Charbonneau2017}.
As illustrated in Fig.~\ref{fig:schematic}a, this approach predicts that on 
increasing the pressure, 
the equilibrium liquid line reaches a dynamical transition point, 
and then terminates at an ideal glass transition critical point, 
just like in the Random First Order Transition (RFOT) scenario first 
introduced by Kirkpatrick, Thirumalai and Wolynes~\cite{kirkpatrick1989scaling} 
and later developed 
by Wolynes and collaborators~\cite{xia2000fragilities,lubchenko2007theory}.
At higher pressure, the mean field approach predicts a glass transition line 
that ends in the infinite-pressure 
limit at a jamming transition point. Here the gap between neighbour particles 
vanishes with a critical exponent, 
as the pressure diverges.
Surprisingly this and other critical exponents are found
~\cite{charbonneau2014} 
to be consistent with those found numerically in finite 
dimensions~\cite{ohernpre,GLN2012}.

One important question is whether this overall scenario survives in finite 
dimensions.
One possibility is that in finite dimensions this scenario disappears, the glass 
transition being a purely kinetic dynamic transition with no diverging static correlation 
length~\cite{Garrahan,Hedges}.
Alternatevely, according to RFOT, the dynamical transition~\cite{gotze1991liquids,gotze2009,Maimbourg} becomes a 
crossover towards the ideal glass transition, where a static critical length associated 
to the cooperative rearranging regions~\cite{kirkpatrick1989scaling,xia2000fragilities,lubchenko2007theory}, 
as originally introduced by Adam and Gibbs~\cite{AG}, diverges.
However, in this scenario it is not clear~\cite{kurchan2009,Liu10} whether the 
ideal glass transition and the jamming transition would coincide~\cite{Liu07} or 
not~\cite{Berthier09,Berthier16} in finite dimension.

In this paper, we investigate the relation between the glass and the jamming 
transitions in finite dimensions,
extending the Cell Theory of the Glass Transition~\cite{AsCo04}, which 
reproduces the essential features 
of Free Volume Theory~\cite{cohen1961,cohen1959,cohen1969}. 
Working within the RFOT scenario, we extend the cell theory taking into account 
the existence of cooperatively rearranging 
regions~\cite{kirkpatrick1989scaling,xia2000fragilities,lubchenko2007theory}. 
As a consequence, local properties like the free volume distribution are 
modified due to the presence of the cooperative length. Starting from this free volume 
distribution, we give arguments indicating that for hard sphere systems
the cooperative length diverges at the jamming transition density. 
As a consequence, 
as illustrated in Fig.~\ref{fig:schematic}b,
the ideal glass critical density $\rho_K$ coincides 
with the glass close packing density~\cite{PZ2010},
or {\it ideal} jamming density $\rho_j$, 
where the cooperative length diverges as $\xi \sim (\rho_j-\rho)^{-\nu}$, with 
$\nu = 2/d$ according to the RFOT.
The experimental jamming transition realized via an out-of-equilibrium 
protocols~\cite{Berthier09,Berthier16,Chaudhuri10,Ciamarra10}
is expected to occur at a protocol dependent volume fraction bounded by 
$\rho_j$, as also illustrated in the figure.
The identification of the glass and of the jamming transitions is supported by 
the fact that $\xi$ results proportional
to the hyperuniform length, considering that hyperuniformity
has been proposed as an essential property of the 
maximum jamming 
density~\cite{torquato2003local,zachary2011hyperuniform,
hopkins2012nonequilibrium,atkinson}.

We checked the prediction for $\nu$ via large-scale 
simulations using finite size scaling at the 
jamming transition, where we find numerical values in $2d$ and $3d$ in excellent 
agreement with the predicted exponent $\nu = 2/d$.
This critical point can be described as a mixed order percolation transition, 
with the order parameter jumping discontinuously at the transition, $\beta = 0$, 
the mean cluster size diverging with exponent $\gamma =2$,
and the number of clusters of size $\xi$ vanishing with an exponent $2-\alpha 
=2$ independent on the dimensionality.
We name this percolative description of the RFOT theory Random First Order 
Percolation Transition, RFOPT.
This scenario is quite different from other mixed-order percolation transition, 
such as the boothstrap percolation~\cite{chalupa,SLC,sellitto,DFC}.
The hallmark of RFOPT is the presence of critical exponents that do not depend 
on the dimensionality, 
which is a typical property of the jamming transition.

\section{Cell Theory}
We start by shortly reviewing the Cell Theory of the Glass 
transition~\cite{AsCo03b,AsCo03,AsCo04},
that combines lattice theories of liquids~\cite{hill1956statistical}
and the ideas of inherent structures, 
free-volume theory, and geometrical packing properties 
to understand the complex dynamics of glass-forming liquids, granular packings, 
and amorphous solids.
In the lattice theory of liquids~\cite{hill1956statistical}, 
the system is divided in $N$ identical cells each corresponding to the unitary 
cell of the underlying crystal. 
Analogously, in the cell theory of glass--forming liquids, 
the system is divided in $N$ Voronoi cells corresponding to the generic 
underlying inherent state,
but one also considers that there are $\Omega$
statistically independent inherent states. 
Consequently, the partition function of a system of $N$ hard sphere particles in 
a volume $V$ is:
\begin{equation}\label{FendN1}
Z = \sum_{ \{N({\mathbf n})\} }\Omega(\{N({\mathbf n})\})
e^{ -\beta F(\{ N({\mathbf n}) \}) },
\end{equation}
where
\begin{equation}\label{FalphaN1}
F(\{  N({\mathbf n})  \})
=
- kT\sum_{\mathbf n} N({\mathbf n})
 \big[\ln \frac{v_f({\mathbf n})}{\Lambda^d\mathcal P({\mathbf n})}\big]
,
\end{equation}
is the  free energy of the inherent states~\cite{AsCo03b,AsCo03,AsCo04}, 
$\Lambda$ is the de Broglie length, 
${\mathbf n}$ is a discrete index referring to the properties of the cell,
\textcolor{black}{$N({\mathbf n})$ the number of cells sharing the same index ${\mathbf n}$,}
$v_f({\mathbf n})$ the \emph{`free volume'} associated with a particle in a cell~\cite{hill1956statistical}.
${\mathcal P}({\mathbf n})$ is the probability to find  a cell with ${\mathbf n}$ that is single-occupied.
\textcolor{black}{The free volume~\cite{hill1956statistical} of a particle is the volume the particle can explore,
averaged over all possible configurations of the other particles, 
in a system restrained to single occupancy of cells.}
The quantity $\Omega(\{N({\mathbf n})\})$
in Eq.~\ref{FendN1} counts the number of distinct space--partitions
(associated with the inherent states) made with the same set of
$\{N({\mathbf n})\}$.
The key elements to be estimated in Eq.~\ref{FalphaN1} are ${\mathcal 
P}({\mathbf n})$ and $\Omega(\{N({\mathbf n})\})$.
At high density, which will be always considered throughout the paper, 
all Voronoi cells of the underlying inherent structure 
are essentially singly occupied and ${\mathcal P}({\mathbf n}) \sim 1$.
$\Omega(\{N({\mathbf n})\})$ can be estimated~\cite{AsCo04,AsCo03,AsCo03b} as 
the number of distinct
configurations that can be made by positioning in
different ways the $N$ cells distributed in groups of  $N({\mathbf
n})$, namely  $\Omega(\{N({\mathbf n})\})=N! /{ \prod_{\mathbf n}N({\mathbf n})! 
}$.
The partition function, Eq.~\ref{FendN1}, can be calculated via a saddle-point 
approximation,
where the sum over all the distributions $\{N({\mathbf n})\}$ is
replaced with the contribution from a distribution $N^*({\mathbf n})$ which 
minimizes the total free energy $\ln Z$.
From this~\cite{AsCo04,AsCo03,AsCo03b}, it is possible to derive the 
distribution of the free volumes, that in the continuum limit becomes
$p(v_f) =\frac{4}{\Gamma(2)}\frac{v_f}{\left< v_f\right>^{2}}\exp{\Big( 
-2\frac{v_f}{\left< v_f \right>} \Big)}$,
where $\left<v_f\right>$ is the average free volume per particle and $\Gamma(.)$ 
is the Gamma function.
The distribution essentially coincides with the distribution originally 
evaluated
within the free volume theory~\cite{cohen1961,cohen1959,cohen1969}, which 
predicts a simple exponential form.

\section{Extended Theory}
The cell theory can be extended to take into account the presence of 
the cooperatively rearranging regions of 
RFOT~\cite{kirkpatrick1989scaling,xia2000fragilities,lubchenko2007theory},
one could identify through a set-to-point correlation 
approach~\cite{verrocchio}.
Within RFOT the system is partitioned in droplets of linear size $\xi$
that, as in first-order transitions, have a free energy of nucleation
containing a volume term plus a surface term.
\textcolor{black}{In standard nucleation theory the surface
term is proportional to $\xi^{d-1}$, while in RFOT it is argued
that} the surface term is proportional to $\xi^{\theta}$  
with $\theta = d/2$, $d$ being the space dimension
(see also~\cite{bouchaud2004adam,berthier2011theoretical} for further 
elaboration and a discussion of other possible values of $\theta$).
The theory predicts that the configurational entropy $s_c $ vanishes as the size 
of the droplet 
diverges $s_c \sim 1/\xi ^{d-\theta}$, while the relaxation time diverges 
exponentially.
\textcolor{black}{Since the total entropy inside the droplet is $\xi^d s_c \simeq \xi^\theta$,
according to RFOT the number of effective degree of freedom of a droplet scales as $\xi^\theta$.
The overall number of degree of freedom, which equals the number of droplets times the degree of freedom
per droplet, is thus reduced by a 
}
factor
$\lambda = (\xi/r_0) ^{d-\theta}$, 
with $r_0$ a characteristic size such that $\rho r_0^d =1$, with $\rho$ being 
the  particle density.
Within the cell theory, this leads to a reduction in the number of 
configurations 
$\Omega$, appearing in the partition function, Eq.~\ref{FendN1}, which now 
becomes
$\Omega (\{N({\mathbf n})\}) = \frac{ (N/\lambda)! }{ \prod_{\mathbf 
n}(N({\mathbf n})/\lambda)! }$,
\textcolor{black}{where $(N({\mathbf n})/\lambda)$
physically represents the number of particles which are able to move characterized by the same cell index ${\mathbf n}$. 
} Using Stirling's approximation:
\begin{equation}\label{sc1}
\ln\Omega (\{N({\mathbf n})\})
\simeq
- \sum_{\mathbf n} \frac{N({\mathbf n})}{\lambda} \ln
\frac{N({\mathbf n})}{N}.
\end{equation}
Following the same procedure as for the case $\lambda=1$ 
we derive a new free volume distribution
\begin{equation}\label{N*2}
p(v_f) =
\frac{k^{k}}{\Gamma(k)}\frac{v_f^{k-1}}{\left< v_f\right>^{k}}
\exp{\Big( -k \frac{v_f}{\left< v_f \right>} \Big)},
\end{equation}
with~\cite{AsteKGammaPRE08} $k=1 + \lambda$.
From  Eq.~\ref{FendN1}, Eq.~\ref{sc1} and Eq.~\ref{N*2} we find:
\begin{equation}\label{lnZ}
\frac{\ln Z}{N}
=
\int p(v_f)
\ln \frac{v_f}{\Lambda^d} dv_f +s_c,
\end{equation}
where $s_c = -\frac{B}{\lambda}$
is the configurational entropy, Eq.~\ref{sc1}, with $B = \int p(v_f)\ln 
p(v_f)dv_f$ 
being a smooth function of $\left<v_f\right>$. 
Since  $\lambda = (\xi/r_0) ^{d-\theta}$, we recover $\xi \propto 
s_c^\frac{1}{\theta-d}$ as in RFOT. 
\textcolor{black}{For molecular liquids Kauzmann assumed~\cite{Kauzmann} , $s_c \propto (T - T_k)$.
In hard sphere systems, where the control parameter is the density and not the temperature,
one assumes $s_c \propto (\rho_K - \rho)$. With this assumption, the critical behaviour of the correlation length is $\xi 
\propto (\rho_K-\rho)^{-1/(d-\theta)}$ and, for $\theta = d/2$, $\nu = 2/d$.}

\textcolor{black}{
The free volume distribution Eq.~\ref{N*2} tends to a delta function in the limit
$k \to \infty$, i.e. $s_c \to 0$. In this limit, $\xi \to \infty$ and $\rho \to \rho_k$.
Accordingly, if $\xi$ diverges for all values of the density between $\rho_k$ and $\rho_j$,
then the free volume distribution is a delta function in this density interval.
This scenario is in princple possible, also considering how the free volume is defined within
the cell theory of liquids, but it is contrary to the 
expectation that, due to the disorder of the system,
only at jamming the free volume distribution would be a delta function.
We therefore argue that the volume distribution is a delta function only at jamming.
This assumption implies $\rho_j = \rho_K$, and
}
\begin{equation}
\label{eq:xi}
\xi/r_0 \sim (\rho_j -\rho)^{-2/(d-\theta)}.
\end{equation} 
We stress that the coincidence of the jamming transition and the glass transition does not depend on the assumed value of $\theta$,
which is an important point as the value of $\theta$ is controversial and of difficult experimental determination.
Note that, since $p(v_f)$ approaches a $\delta$-function for $\left< v_f \right> \to 
0$, our approach predicts that the ideal jammed configuration has no rattlers,
\textcolor{black}{as in mean field~\cite{Charbonneau2012}.}

From Eq.~\ref{eq:xi} one determines the dependence of $k$ on $\rho$ and therefore, 
assuming~\cite{sastry} $\left<v_f\right>^{1/x} = A(\rho_j-\rho)$ where $A$ is a 
constant, the dependence of $k$ on $\left<v_f\right>$. This allows to derive the equation 
of state from Eq.~\ref{N*2} and Eq.~\ref{lnZ}, $P=k_BT\Big(\partial\ln Z/\partial 
V\Big)_{N,T}$,
\begin{equation}\label{pressure2}
\frac{P}{Nk_bT}=\frac{x}{V-V_J}.
\end{equation}
Here $\rho_j=N/V_j $ is the jamming density corresponding to infinite pressure. 
Free volume theory~\cite{wood} predicted Eq.~\ref{pressure2} with $x = d$.

From Eq.~\ref{pressure2} we find the compressibility
to vanish at the jamming transition as $\kappa_T\sim 
\left(\rho_j-\rho\right)^2$.
Using Eq.~\ref{eq:xi} it follows that the relation between the compressibility 
and the cooperative length 
is given by $\kappa_T\sim  (\xi/r_0)^{-d}$.
Interestingly the vanishing of the compressibility in monodisperse jammed 
particles has been linked 
to the concept of hyperuniformity. 
More precisely close to the jamming glass state in $3d$ it was found 
$\kappa_T\sim  (\xi_{DCF}) ^{-3}$, 
where $\xi_{DCF}$,  is a diverging length defined through to the direct pair 
correlation 
function~\cite{torquato2003local,zachary2011hyperuniform,
hopkins2012nonequilibrium}.
This implies that in 3d the hyperuniform length and the cooperative length are 
proportional close to the jamming glass transition $\xi \sim \xi_{DCF}$.
While the link between jamming and hyperuniformity has been 
questioned~\cite{Ikeda2015}
our results support the speculation by Atkinson et al.~\cite{atkinson} 
according to which exact hyperuniformity occurs in packings with no rattlers.

\textcolor{black}{
We now examine how the predictions of the extended cell theory compare to available numerical data.
First, we consider the prediction $\rho_k = \rho_j$, and the related
equation of state of Eq.~\ref{pressure2}.
We remind that $\rho_k$ is the ideal-glass transition density identified from the divergence
of the cooperative correlation length $\xi$, while $\rho_j$ is the density at which the pressure diverges, along the metastable fluid branch. 
A precise measurement of both $\rho_k$ and $\rho_j$ is made difficult by the need of equilibrating the system, as the divergence of the correlation length implies the growth of the relaxation time. A numerical investigation of the dependence of the cooperative correlation length on the pressure,
however, found results consistent with the correlation length diverging in the infinite pressure limit,
and thus consistent with the  $\rho_k = \rho_j$ prediction~\cite{Charbonneau2013}. 
Previous numerical investigations
of the equation of state along the equilibrium metastable branch~\cite{Rintoul,Liu07} for monodisperse
hard sphere support the equation of state of Eq.~\ref{pressure2}.
As a side remark, we note that Eq.~\ref{pressure2}
is quite robust, as it is also recovered during the slow decompression of jammed packings
produced with a fast compression, whose jamming density is not the maximal one~\cite{donev,sastry}.
The extrapolated volume fraction at which the pressure diverges along the metastable fluid branch is found to be 
$\phi_j = 0.644\pm0.005$~\cite{Rintoul} and $\phi_j=0.640\pm0.006$~\cite{Liu07},
and consistent with the maximal volume fraction of disordered jammed configurations
prepared via diverse out-of-equilibrium procedure~\cite{Zhang2005,Scott69,Berryman1983,Jodrey,Ciamarra10}.
\
This supports the idea that the pressure
along the metastable equilibrium branch diverges at the upper bound of the j-line, as in Fig.
\ref{fig:schematic}b.
Of course, numerical results are not a proof. In particular, we note that the numerical estimates
of the pressure/density at which the correlation length diverges are notoriously difficult.
In addition, the estimate of $\rho_j$ need to be considered with care as it is difficult to 
equilibrate hard sphere systems at high densities, because partial ordering might intervene.
\\
The other predictions of our theory, namely the form of the free volume distribution 
along the metastable fluid branch, 
and its convergence to a delta function as $\rho \to \rho_k$, which implies the absence of rattlers, 
cannot be directly tested against literature data. 
Indeed, previous investigations considered the free volume distribution of the particles averaging
over different configurations, in the liquid phase~\cite{fv_liquid,maitiepje,kumar}
or in out of equilibrium high density states~\cite{sastry},
while within the cell theory the free volume of a particle
is the average over the configurations of a same inherent structure.
Since these two measurements give different results for crystals, unless the system
is close packed, we expect them to also give different results along the metastable 
liquid branch, unless the system is at the ideal transition, where according to
both definition the free volume distribution is a delta in zero, and the system has no rattlers.
\\
We also notice that in monodisperse systems the fraction of rattlers along the j-line has 
has been found~\cite{sastry} to vary from $1.7\%$ at $\phi = 0.6392$ to $1.6\%$ at $\phi = 0.6419$,
in agreement with the possibility that the number of rattlers vanishes at $\rho_j$.
However, a detailed investigation of this issue should be carried out.
In this respect, we remind that jammed configurations are produced through out-of-equilibrium 
protocols, so that the fraction of rattlers is also protocol dependent~\cite{Torquato2000}. 
For instance, for large system sizes the fraction of rattlers of jammed packings
of hard-sphere systems converges towards 0.025-0.030, 
for packings prepared using the Lubachevsky-Stillinger algorithm,
and towards 0.015, for packings prepared using the Torquato-Jiao protocol~\cite{Atkinson2013}. 
}

\section{Random First Order Percolation Transition}
\begin{figure}[t!]
\includegraphics*[scale=0.33]{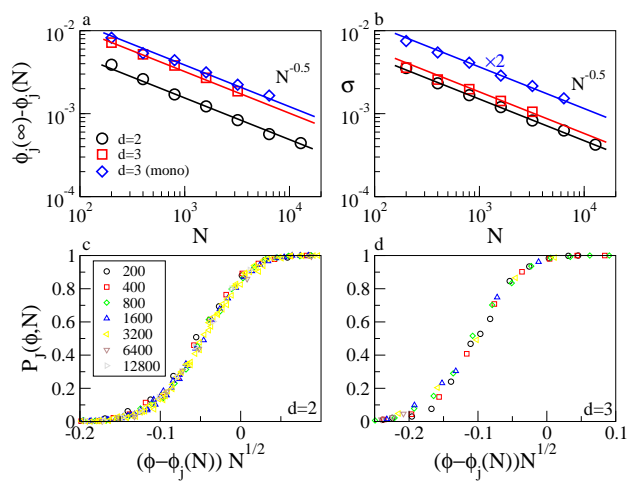}
\caption{
Finite size scaling of the jamming probabilities of soft particles interacting 
via an Harmonic potential,
in two and three dimensions. We show data for a mixture of particles with size 
ratio 1.4 and, in three dimensions,
also for a monodisperse system.
The jamming probability is computed as the average over $10^2-10^3$ independent 
configurations
for every system size $N$ and volume fraction $\phi = \rho v_p$, with $v_p$ 
average particle volume.
The configurations are prepared minimizing the energy with the CG protocol~\cite{ohernprl,ohernpre}.
The distributions are well approximated by an error function, from which
we extract a size dependent jamming volume fraction, $\phi_J(N)$, and a width, 
$\sigma(N)$.
We find $\phi_J(\infty)-\phi_J(N) \propto N^{-1/2}$, where $\phi_J(\infty)$ is 
the system and protocol dependent
jamming volume fraction in the thermodynamic limit, as illustrated in panel a,
and $\sigma(N)\propto N^{-1/2}$, as illustrated in panel b.
Panels c and d illustrate the collapse of the jamming probability distributions.
Analogous results are obtained for monodisperse systems, in three dimensions, as 
well
using a different preparation protocol~\cite{Ciamarra10} to explore the J-line.
\label{fig:fss}
}
\end{figure}

The above results allow to interpret the jamming glass transition
as a percolation transition, we name Random First Order Percolation Transition.
This is a percolation transition 
of compact clusters~\cite{verrocchio} of linear dimension $\xi$ and fractal 
dimension $D=d$~\cite{Stevenson2006,Tarjus2005,Albert2016}.
The size $s^*$ of the critical cluster is given by $s^* \sim \xi^d$. 
As the transition is approached the number density of critical clusters vanishes as 
$\xi^{-d}$,
and the percolation probability jumps discontinuously from $0$ to $1$.
Adapting the standard scaling ansatz~\cite{stauffer,havlin,CF} for the cluster 
size distribution $n(s)$ 
to this peculiar random first-order percolation, we obtain $n(s) \sim 
1/s^{\tau}f(s/s^*)$, 
where $f(x)$ is a rapidly decreasing function for $x \gg 1$, that scales as 
$f(x) \propto x^{-\tau}$ 
with $\tau =d/D +1=2$ for $x \ll 1$.
The number density of clusters scales as $\int n(s) ds \propto \xi^{-D(\tau - 1)} 
\propto (\rho_j -\rho)^{2-\alpha}$.
Thus,  given Eq.~\ref{eq:xi}, and assuming $\theta = d/2$ (see below for the influence of the value $\theta$), we find $\alpha =0$.
The mean cluster size~\cite{stauffer} $S = \sum s^2 n(s)/\sum s n(s)$ diverges at jamming as a power law $S 
\sim (\rho_j -\rho)^{-\gamma}$
with  $\gamma = \nu (d+D)/2 = 2$.
The critical exponent associated to the order parameter is $\beta=\nu(d-D)=0$.
Summarizing, these exponents satisfy the scaling and hyperscaling law:
\begin{equation}
\label{scaling}
2\beta + \gamma = 2-\alpha = d\nu.
\end{equation}
Interestingly we have $\alpha=0$, $\beta=0$, $\gamma = 2$ which are independent 
on the dimensionality.
A similar independence of the critical exponents on the dimensionality 
characterizes the jamming transition.
The only exponent depending on the dimensionality is $\nu$ that equals $2/d$.
This exponent can be estimated via standard
finite-size scaling of the jamming transition density, $\Delta\rho(N) = 
\rho_j^*-\rho_j(N)$,
where $\rho_j^*$ is the estimated jamming density in the thermodynamic limit,
and of the width of the jamming probability distribution, $\sigma(N)$,
that scale as $\Delta \rho \propto \sigma \propto N^{-1/{d\nu}}$.
It is in principle impossible to check this prediction, as it 
refers to the equilibrium jamming point
which is not accessible in the simulations. However, one might expect the 
critical exponent to be the same for all
jamming volume fractions.
Literature investigations of these scaling 
relations~\cite{ohernprl,ohernpre,Ciamarra10,Liu10}
conducted at the smallest value of the jamming volume fraction 
along the J--line
do not unambiguously fix the value of $\nu$~\cite{notenu}.
We have evaluated the probability distribution of the jamming thresholds through 
large-scale simulations,
using different jamming protocols to access different volume 
fractions along the J--line.
Our results, reported in Fig.~\ref{fig:fss},
show that $\Delta \rho \propto \sigma \propto N^{-1/2}$ in both $2d$ and $3d$,
consistently with our theoretical prediction, $\nu =2/d$.

We stress that this scenario refers to the transition as approached from the 
unjammed phase,
the other side of the transition being not accessible in hard-sphere systems.
In soft sphere systems, one might observe different geometrical exponents above 
the transition
should the jammed and the unjammed phase be separated by a 
singularity~\cite{Silbert2005}.
In addition, above the transition other exponents, satisfying their own scaling 
relations, describe the elastic response of the system~\cite{liu2016}.

The percolation exponents depend on the value of $\theta$, in particular for a generic value of $\theta$ 
one finds $\alpha = 2-d/(d-\theta)$, $\gamma=d/(d-\theta)$, $\nu= 1/(d-\theta)$. 
However the exponents still satisfy the scaling and hyperscaling relation of Eq.~\ref{scaling}. 
In any case the value $\nu =2/d$  suggested by the finite size scaling reported in Fig.~\ref{fig:fss}
are consistent with $\theta = d/2$ as assumed. 

\section{Conclusions}
In conclusion using a cell theory previously developed, 
combined with the RFOT approach, we have suggested
that for a monodisperse hard sphere system in finite dimension
the ideal glass transition, where the cooperative length of the RFOT diverges, 
coincides with the ideal jamming transition which occurs at infinite pressure 
at the end of an equilibrium line where the jamming density is maximal without 
rattlers, as in Fig.~\ref{fig:schematic}b.
Following~\cite{AG,xia2000fragilities}, the relaxation times $\tau$ diverges
a la Vogel-Fulcher on approaching the jamming glass critical density, 
$\frac{\tau}{\tau_0} = \exp\left(A/(\rho_j- \rho)\right)$.
Using the properties of the RFOT theory we have been able to describe this glass 
jamming critical point 
as a Random First Order Percolation Transition. 
This allows rationalizing the critical behaviour of the jamming transition in 
terms of an order parameter which 
jumps discontinuously, a critical length with a critical exponent dependent on 
the dimensionality, and all other critical exponents independent on the 
dimensionality. 
These exponents satisfy scaling laws typical of critical phenomena, 
Eq.~\ref{scaling}. 
We note that it is possible to associate other critical exponents to the jamming 
transition.
This is common within percolation theory where, besides usual critical exponents 
obeying standard scaling laws, 
many other critical exponents are introduced, related to quantities, like 
shortest path, 
backbone, elastic properties and so on~\cite{stauffer,havlin,CF}, reflecting 
structural properties of the critical clusters.
\textcolor{black}{Some of the theoretical predictions, such as the coincidence between
$\rho_k$ and $\rho_j$, are supported by numerical results, while others
are currently difficult to test. 
Thus, our paper may stimulate other research to prove or disprove some of the predictions 
we have shown to derive from the unification of the free volume and of the RFOT theory. 
If future results disprove these predictions then some of the basic ingredients
of the proposed approach, such as the free volume theory, the RFOT, or some of our assumptions, must be reconsidered.}
\textcolor{black}{
An interesting open question ahead is the generalization
of this scenario to polydisperse systems, that might satisfy a different
equation of state~\cite{Berthier16} and appears to have rattlers at their maximum jamming density~\cite{Chaudhuri10},
as well as to describe systems of non-spherical particles~\cite{makse}.}
In summary, our work provides insights that change the current theoretical 
interpretation 
of the relation between the glass and the jamming transitions, 
supporting the original suggestion of Ref.~\citenum{nagel1998}, and proposing a 
novel 
percolative interpretation of this transition inspired by RFOT.

\noindent{{\bf Acknowledgement}\\
AC and MPC thank A. de Candia, A. Fierro, R. Pastore, M.Tarzia for discussions, 
and acknowledge from the CNR-NTU joint laboratory Amorphous materials for energy 
harvesting applications.
We also thank D. Leporini, S.Torquato, P. Wolynes 
and L. Berthier for useful comments.
MPC acknowledge support from the Singapore Ministry of Education through the 
Academic Research Fund (Tier 1) 
under Projects No. RG104/15 and RG179/15.


\begin{thebibliography}{29}
\bibitem{notaf}
In standard scaling, $f(x) = const.$ for $x \ll 1$, as the cluster size 
distribution
is a power law at the critical point.
In RFOPT, at the critical point there is a percolating cluster and no finite 
clusters, and therefore $f(x) \propto x^{-\tau}$ for $x \ll 1$.
\bibitem{notenu}
Precisely, in $3d$ results consistent with our predicted scaling $\nu = 2/d$ are 
observed for both $\Delta\rho$ and $\sigma$,
while in $2d$ it is surprisingly found $\Delta \rho \not\propto \sigma$, with 
$\sigma$
exhibiting the expected scaling, and $\Delta \rho \propto N^{-0.7}$.
\end{thebibliography}
\end{document}